\documentclass[11pt]{article}
\usepackage[margin=1in,a4paper]{geometry}
\usepackage{mathtools}
\usepackage{listings}
\usepackage{lstbayes}
\usepackage{xcolor}
\usepackage{graphicx}
\usepackage{endnotes}
\usepackage{authblk}
\usepackage[colorlinks]{hyperref}
\usepackage[doublespacing]{setspace}

\let\footnote=\endnote{}

\title{Exploring Bayesian approaches to eQTL mapping through probabilistic programming}
\author[1,2,3,4]{\small Dimitrios V. Vavoulis}
\date{}

\affil[1]{\footnotesize Department of Oncology, University of Oxford, Oxford, UK}
\affil[2]{\footnotesize The Wellcome Centre for Human Genetics, University of Oxford, Oxford, UK}
\affil[3]{\footnotesize NHS Translational Molecular Diagnostics Centre, Oxford University Hospitals, Oxford, UK}
\affil[4]{\footnotesize NIHR Oxford Biomedical Research Centre, Oxford, UK}
    
\begin{document}
\maketitle
\begin{abstract}
The discovery of genomic polymorphisms
influencing gene expression (also known as \emph{expression quantitative
trait loci} or \emph{eQTLs}) can be formulated as a sparse Bayesian
multivariate/multiple regression problem. An important aspect in the
development of such models is the implementation of bespoke inference
methodologies, a process which can become quite laborious, when multiple
candidate models are being considered. We describe automatic, black-box 
inference in such models using \texttt{Stan}, a popular probabilistic 
programming language. The utilization of systems like \texttt{Stan} can 
facilitate model prototyping and testing, thus accelerating the data 
modelling process. The code described in this chapter can be found at
\href{https://github.com/dvav/eQTLBookChapter}{https://github.com/dvav/eQTLBookChapter}.
\end{abstract} 


\section{Introduction}

The study of genomic variation and its association with gene expression
is critical for elucidating the genetic basis of complex traits,
including diseases. The advent of next-generation sequencing (NGS) made
possible the detailed investigation of this relationship (also known as
\emph{eQTL mapping}) in large cohorts, but it also gave rise to novel
statistical challenges\cite{GTEx_Consortium2013-jf,Lappalainen2013-ex}.\ 
eQTL mapping can be examined in the context of
sparse Bayesian multivariate/multiple regression, where we typically
propose a number of candidate statistical models followed by benchmarking 
them against each other in terms of computational and statistical efficiency\cite{Vavoulis2017-em}.
The most laborious aspect of this process is the development and
software implementation of statistical inference algorithms for each
model under consideration, a task that can be impeded by the potential
absence of favourable mathematical properties (e.g.~conjugacy) in any of
these models.

The aim of this chapter is to demonstrate the utility of a popular
probabilistic programming language (PPL), \texttt{Stan}, in the prototyping and
testing phases of the data modelling process for eQTL mapping\cite{Carpenter2017-gn}. 
PPLs make it possible 
to describe and perform inference in hierarchical probabilistic models in a 
matter of minutes using a small amount of high-level code. Besides \texttt{Stan}, another
popular PPL is \texttt{PyMC}, a Python-based software offering excellent performance,
a wide range of inference algorithms, and the ability to mix these algorithms in the same
inferential cycle for estimating different parts of a given model\cite{Salvatier2016-ic}. 
The reason for choosing 
\texttt{Stan} in this chapter is its simplicity, which stems from the fact that its syntax is 
very close to the mathematical notation already familiar to statisticians.

The practical part of this chapter covers: (a) the acquisition of genomic variation
and gene expression data from online sources, (b) the use of these to simulate artificial
eQTL datasets with known properties, (c) the implementation of statistical models in \texttt{Stan} 
for eQTL mapping, and (d) the estimation of unknown model parameters using the previously simulated 
datasets. In the remaining part of this introduction, we outline the statistical theory, 
which underpins the practical part of this chapter.

\subsection{Theory}

We assume an \(N\times M\) matrix \(Z=\{z_{ij}\} \) of read counts
quantifying the expression of \(N\) transcripts in \(M\) samples, an
\(N\times M\) matrix \(C=\{c_{ij}\} \) of transcript- and sample-specific
normalization factors, and an \(M\times K\) matrix \(X=\{x_{jk}\} \) of
genotypes indicating the number of minor alleles (0, 1 or 2) in each of
\(K\) bi-allelic genomic loci in \(M\) samples. A matrix
\(\tilde{X}=\{\tilde{x}_{jk}\} \) is derived by standardizing each column of
\(X\).

We introduce an \(N\times K\) sparse matrix \(B\) of regression
coefficients with elements \(\beta_{ik}\), which measure the effect of
variant \(k\) on the expression of transcript \(i\). Estimating \(B\) is
the main focus of subsequent analysis. Typically, sparsity is induced on 
\(B\) by imposing appropriate priors. A common sparsity-inducing prior is 
a two-component mixture of the following form:
\[
  \beta_{ik} \sim (1-\pi_{ik})\delta_0 + \pi_{ik}\mathcal{N}(0, \sigma_i^2)
\]
\noindent where \(\delta_0\) is a point-mass distribution centred at 0. This prior
can set \(\beta_{ik}\) exactly at 0, but posterior estimation requires
stochastic exploration of a \(2^{NK}\)-dimensional discrete space. An alternative
approach with obvious computational advantages would be to adopt a
continuous shrinkage prior; for example, the important class of global-local 
shrinkage priors\cite{PolsonScott2012}:
\[
  \beta_{ik} \sim \mathcal{N}\left(0, \eta^2\zeta_{ik}^2\right) \quad
  \eta^2 \sim \mathrm{p}_\eta(\eta^2) \quad
  \zeta^2_{ik} \sim \mathrm{p}_\zeta(\zeta^2_{ik})
\] 

\noindent where \(\zeta_{ik}\) are local (i.e.~gene- and variant-specific)
shrinkage scales, while \(\eta \) is a global scale controlling the
overall shrinkage of \(B\). The shrinkage profile of \(B\) depends on
the form of \(\mathrm{p}_\eta \) and \(\mathrm{p}_\zeta \). Different
choices of these distributions give rise to different shrinkage priors,
but here we shall use the \emph{horseshoe prior}\cite{Carvalho2010-ad} for which
\(\zeta_{ik}\) and \(\eta \) follow standard and scaled half-Cauchy distributions, 
\(\mathcal{C}^+(0, 1)\) and \(\mathcal{C}^+(0, \alpha)\), respectively.

\subsubsection{Normal model}

For the Normal model, we assume that read counts have been normalized,
\(\tilde{z}_{ij}=\frac{z_{ij}}{c_{ij}}\), and log-transformed, \(y_{ij}=\log(\tilde{z}_{ij} + 1)\). 
The model takes the following form:
\[
  y_{ij} \sim \mathcal{N}\left(\beta_{0i}+\sum_k\beta_{ik}\tilde{x}_{jk}, \sigma_i^2\right) \quad
  \beta_{0i} \sim 1 \quad
  \log\sigma^2_i \sim 1 
\]
\[
  \beta_{ik} \sim \mathcal{N}\left(0, \frac{\bar\sigma^2}{NK}\eta^2\zeta_{ik}^2\right) \quad
  \eta \sim \mathcal{C}^+(0, 1) \quad
  \zeta_{ik} \sim \mathcal{C}^+(0, 1)
\]
\noindent where \(\sigma^2_i\) is the variance of gene \(i\) and \(\sum_k\beta_{ik}\tilde{x}_{jk}\) 
is the effect of genomic variation on the baseline expression \(\beta_{0i}\) of gene \(i\). 
We assume that \(M\) is large, so we can afford to impose a flat prior distribution on 
\(\beta_{0i}\) and \(\log\sigma_i\) over the interval (\(-\infty,+\infty \)).  
A more complicated model would include correlations between
different \(y_{ij}\) variables, additional (e.g.~clinical, environmental
and population) co-variates influencing the baseline gene expression
\(\beta_{0i}\), as well as hierarchical priors on \(\beta_{0i}\) and
\(\sigma^2_i\). Notice that the variance of \(\beta_{ik}\) is proportional to 
\(\bar\sigma^2=\left(\frac{\sum_i\sigma_i}{N}\right){^2}\) 
and inversely proportional to the total number of coefficients. The implicit
assumption under this formulation is that the true global scale parameter is \(\xi \sim \mathcal{C}^+\left(0, \frac{\bar\sigma}{\sqrt{NK}}\right) \). 
Finally, in this and the subsequent models, we assume that we can ignore any gene-specific factors 
(e.g.~length) affecting \(c_{ij}\), thus \(c_{ij}\equiv c_j\).

\subsubsection{Negative Binomial model}

The Negative Binomial distribution is immensely popular for modelling over-dispersed RNA-seq 
data\cite{Love2014-yt,McCarthy2012-bh,Vavoulis2015-ot}, but 
the mathematical complexities associated with inference in this model might explain (at least 
partially) the popularity of transformation-based methods, such as \texttt{voom}\cite{Law2014-uc}. 
Here, we examine the following model:  
\[
  z_{ij} \sim \mathcal{NB}(m_{ij}, \phi_i) \quad
  \log m_{ij} = \log c_{ij} + \log L_j + \beta_{0i}+\sum_k\beta_{ik}\tilde{x}_{jk} \quad
  \beta_{0i} \sim 1 \quad \log\phi_i \sim 1
\]
\[
  \beta_{ik} \sim \mathcal{N}\left(0, \frac{\eta^2\zeta_{ik}^2}{NK}\right) \quad
  \eta \sim \mathcal{C}^+(0, 1) \quad
  \zeta_{ik} \sim \mathcal{C}^+(0, 1)
\]
\noindent where \(m_{ij}\) is the gene- and sample-specific mean of the Negative Binomial distribution, 
and \(\phi_i\) is a gene-specific dispersion parameter, such that \(\mathrm{Var}[z_{ij}]=m_{ij}+m^2_{ij}\phi_i\);
\(L_j=\sum_i z_{ij}\) is the total number of reads in sample \(j\). 

\subsubsection{Poisson-LogNormal model}

An alternative approach to work with non-transformed data is to impose a Poisson observational model on top
of the Normal:
\[
z_{ij} \sim \mathcal{P}(m_{ij}) \quad \log m_{ij} = \log c_{ij} + \log L_j+ y_{ij}
\]
\noindent where \(y_{ij}\) serve as latent variables following the Normal model. The Poisson-LogNormal 
model is motivated by (a) the fact that the Negative Binomial model can be thought of as a 
Poisson-Gamma mixture, and (b) replacing the Gamma distribution in the
previous mixture with LogNormal. By invoking the law of total expectation and the law of total variance, 
we can see that \(E[z_{ij}]=c_{ij}L_j e^{\beta_{0i}+\sum_k\beta_{ik}\tilde{x}_{jk} + \sigma_i^2/2}\) 
and \(\mathrm{Var}[z_{ij}]=E[z_{ij}] + E[z_{ij}]{^2}\phi_i\). Hence, the variance has the same form 
as in the Negative Binomial model, with dispersion parameter \(\phi_i=e^{\sigma^2_i}-1\). When 
\(\sigma^2_i=0\), the above model reduces to Poisson.

\section{Materials}

\subsection{Operating system}

\begin{enumerate}
\item
  A working UNIX environment (e.g.~Linux or MacOS X) with a terminal
  emulator running the command shell \texttt{bash}. The work presented in
  this chapter was tested on Ubuntu Linux v18.04.
\end{enumerate}

\subsection{Software}

\begin{enumerate}
\item
  A recent version of the command line tool and library \texttt{curl} for
  transferring data with URLs.
\item
  A recent version of \texttt{vcftools}, a set of tools written in Perl
  and C++ for working with VCF files\cite{Danecek2011-eu}.
\item
  A recent version of \texttt{R}, the free software environment for
  statistical computing\cite{R}. 
\item
  A recent version of \texttt{rstudio}, an integrated development
  environment for working with \texttt{R} and the command line\cite{RStudio}.
\item 
  A recent version of \texttt{rstan}, an \texttt{R} interface for \texttt{stan}.
\item 
  A recent version of \texttt{plyr}, a set of \texttt{R} tools for splitting, 
  modifying and combining data\cite{plyr}.
\item 
  A recent version of \texttt{doMC}, an \texttt{R} package providing a parallel backend for multicore computation. 
\item 
  A recent version of \texttt{cowplot}, an \texttt{R} package for plotting.
\item 
  A recent version of \texttt{tidyverse}, a collection of \texttt{R} packages 
  for data wrangling and plotting.
  \item 
  A recent version of \texttt{reshape2}, an \texttt{R} package for restructuring and aggregating data\cite{reshape2}.
\end{enumerate}
The above \texttt{R} packages can be installed either through the graphical interface provided by \texttt{rstudio}, or through
the \texttt{R} console; for example \texttt{install.packages('tidyverse')}  % chktex 36

\section{Methods}

\subsection{Data acquisition}

\begin{enumerate}
\item 
  Start \texttt{rstudio}

\item
  Create a working directory tree by typing the following commands at
  the \texttt{bash} command prompt\footnote{
    You can access the command prompt through a terminal emulator. From within \texttt{rstudio}, 
    you can create a terminal emulator by going to \texttt{Tools\textgreater{}Terminal\textgreater{}New Terminal}.
  }:
\begin{lstlisting}[language=bash]
mkdir -p eQTLchapter/{data,R,stan}
\end{lstlisting}     

\item
  Make the root of the tree you just created your working directory by
  typing the following at the \texttt{R} console\footnote{
    Instead of steps 2 and 3, you can just create a new \texttt{rstudio} project (\texttt{File\textgreater{}New
    Project\ldots{}}) with equivalent results
  }:
\begin{lstlisting}  
setwd('eQTLchapter')
\end{lstlisting}

\item
  Download genomic variation data from the 1000 Genomes project\cite{1000_Genomes_Project_Consortium2015-qu}. At the
  \texttt{bash} command prompt, type the following\footnote{
    This chunk of code uses \texttt{curl} to stream data from the 1000 Genomes Project and pipe them 
    (in gzipped VCF format) to \texttt{vcftools} for further processing. Only variants in chromosome 7 
    between positions 100K and 200K are retained. In addition, all variants with FILTER other than PASS are
    removed; indels are removed; variants in non-bi-allelic sites are removed; only variants with allele 
    frequency higher than 5\% are retained (i.e. 0.05\textless{}MAF\textless{}0.95). Filtered data are stored as a matrix 
    with elements 0, 1 and 2 indicating the number of alleles in 496 variants across 2504 samples. If \texttt{vcftools} 
    throws an error about not being able to open temporary files, you need to increase the maximum number 
    of open files allowed by your system to more than double the number of samples in the VCF file. At the 
    \texttt{bash} command prompt, type: \texttt{ulimit\ -n\ 5100}. You may need root permissions to run 
    this command.
  }:
\begin{lstlisting}[language=bash,escapechar=;]  
cd eQTLchapter
curl -o - ;\href{ftp://ftp.1000genomes.ebi.ac.uk/vol1/ftp/release/20130502/ALL.chr7.phase3_shapeit2_mvncall_integrated_v5a.20130502.genotypes.vcf.gz}{link}; \
    | vcftools --gzvcf - \
        --chr 7 --from-bp 100000 --to-bp 200000 \
        --remove-filtered-all --remove-indels \
        --min-alleles 2 --max-alleles 2 \
        --maf 0.05 --max-maf 0.95 \
        --012 --out data/chr7
\end{lstlisting}

\item
  Download RNA-seq read count data from the ReCount project\cite{Frazee2011-ko}. 
  At the \texttt{bash} prompt, type the following:
\begin{lstlisting}[language=bash,escapechar=|]  
curl -o data/montpick_count_table.txt |\href{http://bowtie-bio.sourceforge.net/recount/countTables/montpick_count_table.txt}{link}|
curl -o data/montpick_phenodata.txt |\href{http://bowtie-bio.sourceforge.net/recount/phenotypeTables/montpick_phenodata.txt}{link}| 
\end{lstlisting}  
\end{enumerate}

\subsection{Data importing}

\begin{figure}[ht]
  \centering
  \includegraphics{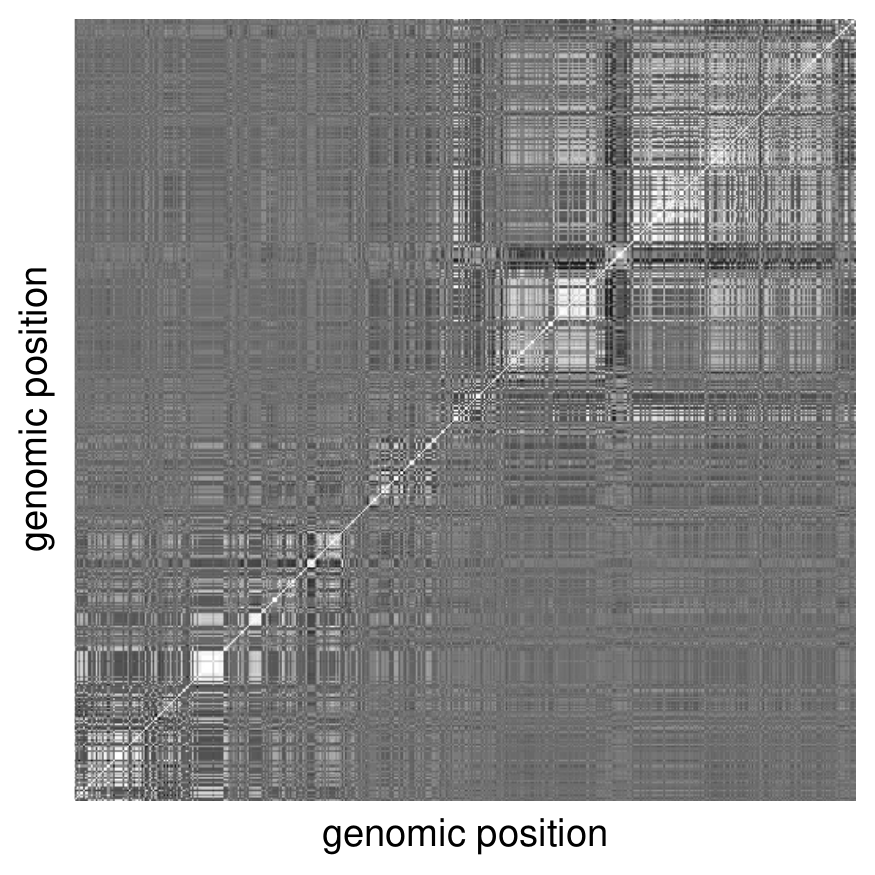}.
  \caption{\label{fig:geno} Correlation structure of the matrix of genotypes. Blocks  
    of highly correlated variants appear in white}
\end{figure}

\begin{enumerate}
\item
  Create the following \texttt{R} scripts. At the \texttt{bash}
  prompt, type:\footnote{
    Alternatively, you can create these files through
    \texttt{rstudio} (\texttt{File\textgreater{}New File\textgreater{}R Script}).
  }:
\begin{lstlisting}[language=bash]  
touch analysis.R R/{utils,viz}.R
\end{lstlisting}

\item
  Using the code editor provided by \texttt{rstudio}, add the line
  \texttt{library(tidyverse)} at the top of \texttt{utils.R} and  % chktex 36
  \texttt{viz.R}, and the following lines at the top of
  \texttt{analysis.R}\footnote{Whenever the code in \texttt{utils.R} or
    \texttt{viz.R} changes, these files have to be sourced again to make
    the changes visible to \texttt{analysis.R} by executing 
    lines 2 and 3 of this code chunk. A more structured approach would be to develop an 
    \texttt{R} package, which is however beyond the scope of this chapter. The function 
    \texttt{doMC::registerDoMC(cores\ =\ 8)}  % chktex 36
    registers a parallel backend with 8 cores. This is required for multicore 
    functionality in the \texttt{plyr} package.}:
\begin{lstlisting}  
source('R/utils.R')
source('R/viz.R')

doMC::registerDoMC(cores=8)
\end{lstlisting}  

\item
  Create a function in \texttt{utils.R} for importing the genotypic data
  into \texttt{R}\footnote{
    In this and all subsequent code listings, we make heavy use of the pipe operator \texttt{\%>\%} to form 
    long linear chains of functions, where the output of each function becomes the input of the next. This allows for the
    generation of readable, easy-to-understand workflows. In this particular function, we read genotypic data from disk files
    in the form of an \(M\times K\) matrix. Before returning, the matrix is filtered by removing all variants (i.e.~columns) that 
    may have zero variance across all samples (lines 24,25). Notice that file names are constructed in a device-independent
    manner using the function \texttt{file.path}. Finally, we prefer the operator \texttt{=} instead of the traditionally used 
    \texttt{<-} for indicating assignment.
  }:
\begin{lstlisting}  
load_genotypes = function() {
  # read sample names
  samples =
    file.path('data', 'chr7.012.indv') %>%
    read_tsv(col_names = 'SAMPLE') %>%
    pull(SAMPLE)

  # read loci
  pos =
    file.path('data', 'chr7.012.pos') %>%
    read_tsv(col_names = c('CHROM', 'POS')) %>%
    unite('POS', CHROM, POS, sep = ':') %>%
    pull(POS)

  # read genotypes
  geno =
    file.path('data', 'chr7.012') %>%
    read_tsv(na = '-1', col_names = c('SAMPLE', pos)) %>%
    select(-1) %>%
    as.matrix()
  dimnames(geno) = list(samples = samples, variants = pos)

  # remove loci with the same genotype across all samples
  sds = apply(geno, 2, sd, na.rm = T)
  geno[,sds > 0]
}
\end{lstlisting}  

\item
  Source \texttt{utils.R}, add the following line to \texttt{analysis.R}
  and execute it in order to import the genotypic data into \texttt{R}:
\begin{lstlisting}  
obs_geno = load_genotypes()
\end{lstlisting}  

\item
  Create the following function in \texttt{viz.R} for visualizing the correlation structure of the
  matrix of genotypes:
\begin{lstlisting}  
plot_genotypes = function(geno) {
  geno %>%
    cor() %>%
    reshape2::melt() %>%
    ggplot() +
    geom_raster(aes(x = Var1, y = Var2, fill = value)) +
    theme(legend.position = 'none',
          axis.text.x = element_blank(),
          axis.text.y = element_blank(),
          axis.ticks.x = element_blank(),
          axis.ticks.y = element_blank()) +
    scale_fill_gradient(low = 'black', high = 'white') +
    labs(x = 'genomic position', y = 'genomic position')
}
\end{lstlisting}  

\item
  Source \texttt{viz.R}, add the following line to \texttt{analysis.R}
  and execute it in order to visualize the genotypic data (Figure~\ref{fig:geno}):
\begin{lstlisting}  
plot_genotypes(obs_geno)
\end{lstlisting}  

\item
  Create a function in \texttt{utils.R} for importing the read counts data
  into \texttt{R}\footnote{
    Data are loaded in the form of an \(N\times M\) matrix of read counts. By default,
    only data for the CEU population are read. If YRI data are preferred, set the function
    argument to \texttt{pop='YRI'}, instead. As for the genotypic data, genes with constant
    expression across all samples are removed (lines 18,19).
  }:
\begin{lstlisting}
load_counts = function(pop = 'CEU') {
  # load samples
  samples =
    read_delim(file.path('data', 'montpick_phenodata.txt'), delim = ' ') %>%
    filter(population == pop) %>%
    pull(sample.id)

  # load read counts
  counts =
    read_tsv(file.path('data', 'montpick_count_table.txt')) %>%
    select(GENE = gene, samples) %>%
    as.data.frame() %>%
    column_to_rownames('GENE') %>%
    as.matrix()
  dimnames(counts) = list(genes = rownames(counts), samples = samples)

  # remove genes with the same number of counts across all samples
  sds = apply(counts, 1, sd, na.rm = T)
  counts[sds > 0,]
}    
\end{lstlisting}

\item
  Source \texttt{utils.R}, add the following line to \texttt{analysis.R}
  and execute it in order to import the read counts data into \texttt{R}:
\begin{lstlisting}  
obs_counts = load_counts()
\end{lstlisting}  
\end{enumerate}

\subsection{Estimation of gene expression statistics}

\begin{figure}[ht]
  \centering
  \includegraphics[width=0.9\textwidth]{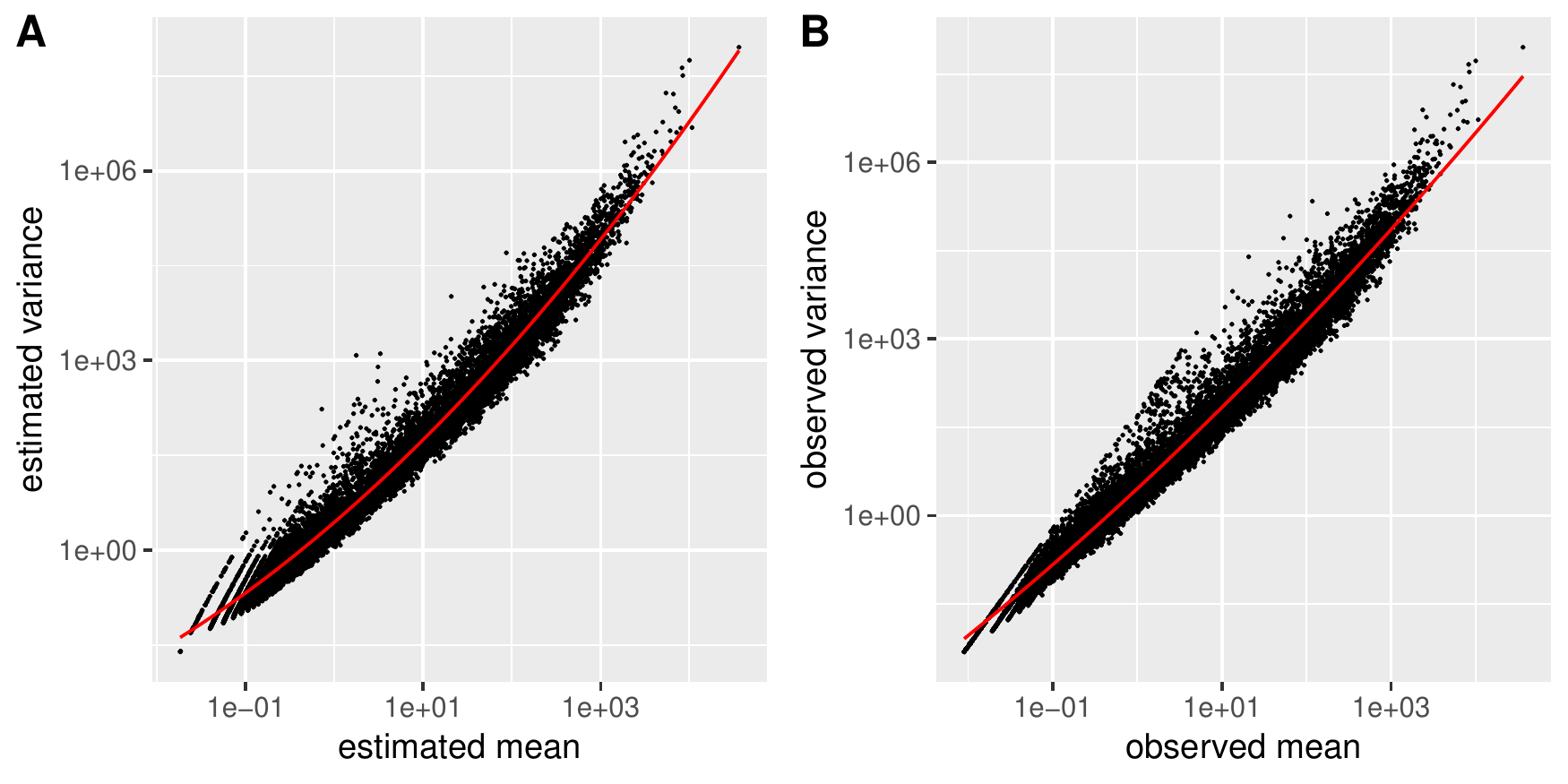}.
  \caption{\label{fig:mean_var} Estimated and observed mean-variance relationship 
    in the gene expression data}
\end{figure}

\begin{enumerate}  
\item
  Create a function in \texttt{utils.R} for calculating sample-specific normalization factors 
  given a matrix of count data\footnote{
    This function implements the \emph{median-of-ratios} method for normalizing a matrix of count data, 
    as presented in~\cite{Love2014-yt}. Calculations are performed on the logarithmic scale.
  }:
\begin{lstlisting}
calculate_norm_factors = function(counts) {
  lcounts = log(counts)
  lgm = rowSums(lcounts) / ncol(lcounts)
  idxs = is.finite(lgm)
  lratios = sweep(lcounts[idxs,], 1, lgm[idxs], '-')
  apply(exp(lratios), 2, median)
}
\end{lstlisting}

\item 
  Create a function in \texttt{utils.R} for calculating the log-likelihood of a vector of counts, assuming each element is sampled 
  from the Negative Binomial distribution\footnote{
    The probability mass function of the Negative Binomial distribution assumed here is parameterized in terms of 
    gene-specific mean \(\mu_i\) and inverse dispersion \(\alpha_i\) parameters, as follows:
    \[
      z_{ij} \sim \frac{\Gamma(z_{ij} + \alpha_i)}{\Gamma(z_{ij}+1)\Gamma(\alpha_i)}
        {\left(\frac{c_j\mu_i}{c_j\mu_i+\alpha_i}\right)}^{z_{ij}}
        {\left(\frac{\alpha_i}{c_j\mu_i+\alpha_i}\right)}^{\alpha_i}
    \]
    \noindent where \(c_j\) are sample-specific normalisation factors.
  }:
\begin{lstlisting}
lognbinom = function(pars, z, cc) {
  mu = pars[1]
  phi = pars[2]

  m = cc * mu
  alpha = 1 / phi
  ll = lgamma(z + alpha) - lgamma(alpha) - lgamma(z + 1) +
    z * log(m) + alpha * log(alpha) - (z + alpha) * log(m + alpha)

  sum(ll)
}  
\end{lstlisting}

\item
  Create a function in \texttt{utils.R} for calculating gene-specific mean and dispersion maximum likelihood 
  estimates for the Negative Binomial distribution given a matrix of count data\footnote{
    Given an \(N\times M\) matrix of counts, we fit the Negative Binomial distribution to each gene/row, and we estimate
    gene-specific maximum likelihood estimates of mean and dispersion parameters. For parameter estimation, we use the function
    \texttt{optim}. The code \texttt{possibly(optim, otherwise = NULL)} on line 3 returns a version of the \texttt{optim} function, % chktex 36
    which returns \texttt{NULL} on error instead of halting execution, thus allowing processing of the whole set of available genes. 
    The parameter \texttt{fnscale=-1} on line 11 signals \texttt{optim} to perform maximization of the log-likelihood 
    \texttt{lognbinom}, instead of minimization. Iterating over the rows of the data matrix is done using the \texttt{plyr} 
    function \texttt{alply}, and it is performed in parallel (argument \texttt{.parallel=T} on line 15). After all rows have 
    been processed, genes for which estimation failed or did not converge are discarded 
    (lines 16,17), results are extracted and properly formatted (lines 18--21), and gene-specific variance values are calculated
    given the estimated values for mean and dispersion parameters (line 22).                  
  }: 
\begin{lstlisting}  
calculate_gene_stats = function(counts) {
  sizes = calculate_norm_factors(counts)
  optim = possibly(optim, otherwise = NULL)
  stats =
    counts %>%
    plyr::alply(1, function(cnts) {
      optim(par = c(1, 1),
            fn = lognbinom,
            z = cnts,
            cc = sizes,
            control = list(fnscale = -1),
            method = 'L-BFGS-B',
            hessian = T,
            lower = 1e-12)
    }, .parallel = T, .dims = T) %>%
    compact() %>%
    discard(~.x$convergence > 0) %>%
    map('par') %>%
    enframe() %>%
    mutate(value = map(value, str_c, collapse = ',')) %>%
    separate(value, into = c('MEAN', 'PHI'), sep = ',', convert = T) %>%
    mutate(VAR = MEAN + PHI * MEAN^2) %>%
    rename(genes = name) %>%
    as.data.frame() %>%
    column_to_rownames('genes') %>%
    as.matrix()
  dimnames(stats) = list(genes = rownames(stats), statistics = colnames(stats))
  stats
}  
\end{lstlisting}

\item
  Source \texttt{utils.R}, add the following line to \texttt{analysis.R}
  and execute it in order to calculate basic gene-wise statistics from the read counts data:
\begin{lstlisting}  
obs_counts_stats = calculate_gene_stats(obs_counts)
\end{lstlisting}  

\item   
  Create a function in \texttt{viz.R} for visualising empirical and estimated gene-wise statistics\footnote{
    An inner function \texttt{plot\_fn} is defined at the top of \texttt{plot\_mean\_variance}, in order to 
    avoid code repetition when generating plots \texttt{gg1} and \texttt{gg2} below.  
  }:
\begin{lstlisting}
plot_mean_variance = function(counts, stats) {
  plot_fn = function(df, xlabel, ylabel) {
    df %>%
      ggplot() +
      geom_point(aes(x = MEAN, y = VAR), size = 0.1) +
      geom_smooth(aes(x = MEAN, y = VAR), size = 0.5,
                  color = 'red', method = 'lm', formula = y ~ poly(x, 2)) +
      scale_x_continuous(trans = 'log10') +
      scale_y_continuous(trans = 'log10') +
      labs(x = xlabel, y = ylabel)
  }

  # normalise count data
  sizes = calculate_norm_factors(counts)
  counts = sweep(counts, 2, sizes, '/')

  # plot of estimated stats
  gg1 =
    stats %>%
    as.data.frame() %>%
    plot_fn(xlabel = 'estimated mean', ylabel = 'estimated variance')

  # plot of observed stats
  gg2 =
    data.frame(
      MEAN = rowMeans(counts),
      VAR = apply(counts, 1, var)
    ) %>%
    plot_fn(xlabel = 'observed mean', ylabel = 'observed variance')

  # combine plots
  cowplot::plot_grid(gg1, gg2, align = 'vh', labels = 'AUTO')
}  
\end{lstlisting}
  
\item
  Source \texttt{viz.R}, add the following line to \texttt{analysis.R}
  and execute it in order to visualise estimated and observed gene-wise 
  statistics (Figure~\ref{fig:mean_var}):
\begin{lstlisting}  
plot_mean_variance(obs_counts, obs_counts_stats)
\end{lstlisting}
\end{enumerate}

\subsection{Data simulation}
\begin{enumerate}
\item
  Create a function in \texttt{utils.R} for simulating genotypic and count gene expression data
  given observed data of the same kind and their estimated statistics\footnote{
    This function should be treated as a starting point for more sophisticated data simulation methods.
    Genotypic data are generated by randomly sampling without replacement \texttt{nsamples} rows and \texttt{nvars} 
    columns from the observed matrix of genotypes. This approach breaks the spatial correlation between variants,
    so the variants/columns in the generated matrix \(X\) can be treated as being mostly uncorrelated. The matrix \(B\)
    of coefficients is initially generated as a matrix of zeros. In a second stage, \texttt{nhits} elements are randomly 
    chosen, and set to non-zero values sampled randomly from a shifted exponential distribution. Half of the \texttt{nhits}
    elements are asigned positive values and the remaining half are assigned negative values. Given matrices \(X\) and 
    \(B\) and the top \texttt{ngenes} estimated mean and dispersion parameters (where parameter values are first ranked 
    in order of decreasing mean values; see line 20), a matrix \(Z\) of read counts is generated with elements sampled 
    from the Negative Binomial distribution. Alternatively, estimated mean and dispersion parameters can be randomly 
    sampled by uncommenting line 19 (and commenting out line 20).      % chktex 36       
  }:
\begin{lstlisting}  
simulate_data = function(count_stats, genotypes, nsamples = 1000, ngenes = 100, 
                         nvars = 50, nhits = 10, rate = 4) {
  # fetch genotypes
  X = genotypes[sample(1:nrow(genotypes), nsamples), 
                sample(1:ncol(genotypes), nvars)]
  X = X[,apply(X, 2, sd, na.rm = T) > 0]
  dimnames(X) = list(samples = str_c('S', 1:nrow(X)), 
                     variants = str_c('V', 1:ncol(X)))

  # simulate matrix of coefficients
  B = matrix(0, nrow = ngenes, ncol = ncol(X), 
              dimnames = list(genes = str_c('G', 1:ngenes),
                              variants = str_c('V', 1:ncol(X))))
  hits = c(1 + rexp(0.5 * nhits, rate = rate), 
          -1 - rexp(0.5 * nhits, rate = rate))
  B[sample(length(B), length(hits))] = hits

  # simulate counts
  # df = count_stats[sample(1:nrow(count_stats), ngenes),]
  df = count_stats[order(count_stats[,'MEAN'], decreasing = T),][1:ngenes,]

  X0 = scale(X, center = T, scale = T)
  m = sweep(exp(B %*% t(X0)), 1, df[,'MEAN'], '*')
  alpha = 1 / df[,'PHI']
  Z = matrix(rnbinom(length(m), mu = m, size = alpha), nrow = nrow(m))
  dimnames(Z) = list(genes = str_c('G', 1:nrow(Z)), 
                     samples = str_c('S', 1:ncol(Z)))

  # output
  lst(B, X, Z, stats = df)
}
\end{lstlisting}
  
\item 
  Source \texttt{utils.R} to make \texttt{simulate\_data} visible to the global environment.
\end{enumerate}

\subsection{Model implementation}
\begin{enumerate}
\item
  Copy the files implementing the Normal, Poisson and Negative Binomial models 
  to the \texttt{stan/} directory (\texttt{normal.stan}, \texttt{poissonln.stan} and \texttt{negbinom.stan}, 
  respectively; see end of chapter for code listings)\footnote{
    The \texttt{stan} code presented here is divided in a number of blocks. The \texttt{data\{\ldots \}}
    block is read exactly once at the beginning of the inference procedure, and it declares required model 
    data. The \texttt{transformed data\{\ldots \}} block in \texttt{negbinom.stan} and \texttt{poissonln.stan} 
    defines transformations 
    of previously declared data, and it is also executed only once at initialisation. The 
    \texttt{parameters\{\ldots \}} block declares model parameters, an unconstrained version of which  
    will be the target of inference procedures. The 
    \texttt{transformed parameters\{\ldots \}} block defines variables in terms of previously defined 
    parameters. All variables that appear in these last two blocks will be returned at the output after 
    inference completes. Finally,
    the \texttt{model\{\ldots \}} block is where the joint log probability function is defined, using
    a syntax very close to the actual mathematical notation. Sampling statements (priors) for all variables 
    in the \texttt{parameters\{\ldots \}} block must appear here, otherwise a uniform, possibly unconstrained, prior 
    is assumed.  
  }.
\item 
  Compile the models by adding the following lines in \texttt{analysis.R} and executing them\footnote{
    The argument \texttt{auto\_write=T} ensures that unecessary re-compilations of \texttt{stan} code 
    are avoided. For this to work, the variables \texttt{negbinom}, \texttt{poissonln} and \texttt{normal}
    must reside in the global environment, as in this code listing.
  }:
\begin{lstlisting}  
normal = 
  file.path('stan', 'normal.stan') %>%
  rstan::stan_model(auto_write = T)
poissonln = 
  file.path('stan', 'poissonln.stan') %>%
  rstan::stan_model(auto_write = T)
negbinom = 
  file.path('stan', 'negbinom.stan') %>%
  rstan::stan_model(auto_write = T)
\end{lstlisting} 
\end{enumerate}

\subsection{Model testing}
\begin{figure}[ht]
  \centering
  \includegraphics[width=0.9\textwidth]{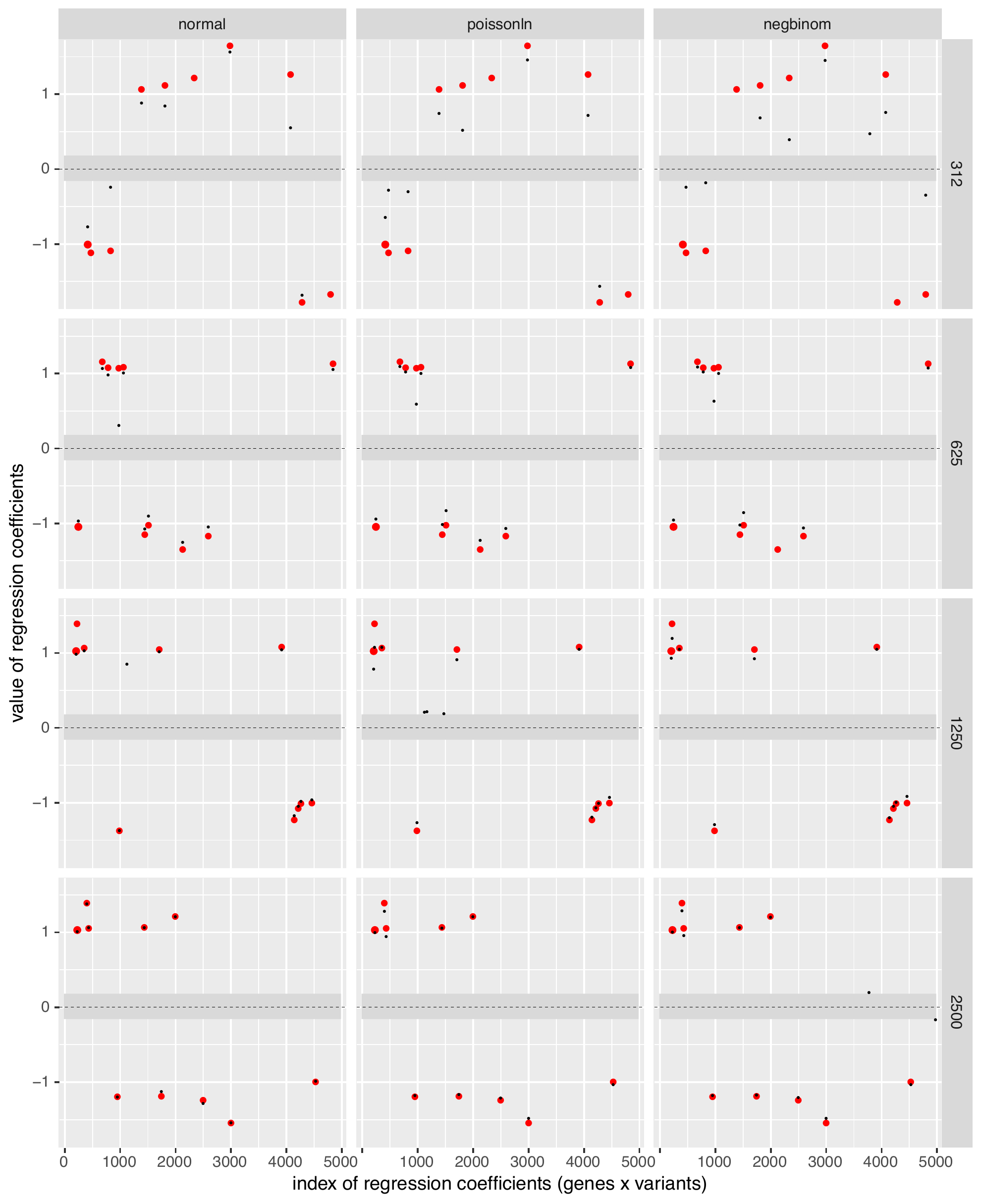}.
  \caption{\label{fig:fitted_models} True (red dots) and estimated (small black dots) 
    regression coefficients for the Normal, Poisson-LogNormal and Negative Binomial models at 
    four different sample sizes. Regression coefficients within the gray band around zero are
    considered of negligible size and they are omitted from the plot.}
\end{figure}

\begin{enumerate}
\item
  Create an auxilliary function in \texttt{utils.R} for fitting a compiled model and extracting the estimated 
  regression coefficients \(B\)\footnote{
    \texttt{Stan} provides three different inference methodologies\cite{StanManual}: (a) Full Bayesian inference using a self-adjusting Hamiltonian
    Monte Carlo approach, (b) approximate Bayesian inference using Automatic Differentiation Variational Inference (ADVI), and (c)
    point parameter estimates by maximizing the joint posterior. Here, for reasons of computational efficiency, we use the third 
    approach (through function \texttt{rstan::optimizing}), which however does not provide any measure of uncertainty of the estimates. 
    In principle, such estimates can be derived at a second stage by approximating locally the posterior distribution using the 
    Laplace approximation. \texttt{rstan::optimizing} takes as input a compiled model and input data in the form of a list of variables,
    as they appear in the \texttt{data\{\ldots \}} block of the model definition (variables not in this block are ignored). Function
    \texttt{fit\_model} returns the matrices of true and estimated coefficients in vector format, and a vector of the corresponding
    indices.    

  }:
\begin{lstlisting}  
fit_model = function(data, model, fcts = calculate_norm_factors(data$Z), ...) {
  Z_tilde = sweep(data$Z, 2, fcts, '/')          # normalise count data
  Y = log(Z_tilde + 1)                           # transform normalised count data
  X0 = scale(data$X, center = T, scale = T)      # standardise genotypes

  # MAP estimation
  fit = rstan::optimizing(object = model,
                          data = list(Z = data$Z,
                                      Y = Y,
                                      X = X0,
                                      c = fcts,
                                      s = colSums(data$Z),
                                      N = nrow(Y),
                                      M = ncol(Y),
                                      K = ncol(X0)),
                          seed = 42,
                          ...)

  # extract estimated matrix of regression coefficients B in vector format
  tibble(EST =
           fit %>%
           pluck('par') %>%
           enframe() %>%
           filter(str_detect(name, '^B\\[')) %>%
           pull(value),
         TRU = as.vector(data$B),
         IDX = 1:length(TRU))
}  
\end{lstlisting} 

\item 
  Source \texttt{utils.R} to make \texttt{fit\_model} visible to the global environment.

\item 
  Make the function \texttt{cpp\_object\_initializer} visible to the global environment by adding 
  the following line to \texttt{analysis.R} and executing it\footnote{
    This is necessary for successfully calling \texttt{rstan::optimizing}. Alternatively, we could have 
    imported the whole of \texttt{rstan} using \texttt{library(rstan)}, and called \texttt{optimizing} % chktex 36
    directly (i.e.~without the \texttt{::} operator).     
  }:
\begin{lstlisting}  
cpp_object_initializer = rstan::cpp_object_initializer
\end{lstlisting} 

\item 
  Set the initial seed of \texttt{R}'s random number generator by adding 
  the following line in \texttt{analysis.R} and executing it\footnote{
    This code facilitates reproducibility, since for the same seed,
    the same sequence of pseudo-random numbers is generated.
  }:
\begin{lstlisting}  
set.seed(42)
\end{lstlisting} 

\item 
  Test all three models on simulated data of various sizes (312, 625, 1250, 2500) by adding the 
  following lines in \texttt{analysis.R} and executing them\footnote{
    This code chunk includes two nested loops: the outer loop iterates over different sample sizes, while the inner
    loop iterates over models, and it is executed in parallel. At each iteration of the outer loop, a dataset of appropriate 
    size is simulated and then passed to each tested model in the inner loop.
  }:
\begin{lstlisting}
fitted_models =
  plyr::ldply(lst(312, 625, 1250, 2500), function(N) {
    sim = simulate_data(obs_counts_stats, obs_geno, nsamples = N)
    plyr::ldply(lst(normal, poissonln, negbinom), function(mdl) {
      fit_model(sim, mdl)
    }, .parallel = T, .id = 'MODEL')
  }, .progress = 'text', .id = 'NSAMPLES') %>%
  as_tibble()
\end{lstlisting}

\item 
  Create a function in \texttt{viz.R} for visualising the results\footnote{
    This function takes a \texttt{thr} argument: estimated coefficient values below this threshold
    are considered effectively 0 and set to NA (line 4).\ When plotting, NA values will be dropped
    with a warning.      
  }:
\begin{lstlisting}
plot_fitted_models = function(df, thr = 0.1 * max(abs(df$EST))) {
  df %>%
    mutate(TRU = na_if(TRU, 0),
           EST = if_else(abs(EST) < thr, NA_real_, EST)) %>%
    ggplot() +
    geom_ribbon(aes(x = IDX, ymin = -thr, ymax = thr), fill = 'grey85') +
    geom_hline(yintercept = 0, linetype = 'dashed', size = 0.2) +
    geom_point(aes(x = IDX, y = TRU), color = 'red') +
    geom_point(aes(x = IDX, y = EST), size = 0.5) +
    facet_grid(NSAMPLES~MODEL) +
    scale_x_continuous(expand = c(0.02, 0.02)) +
    scale_y_continuous(expand = c(0.02, 0.02)) +
    labs(x = 'index of regression coefficients (genes x variants)',
         y = 'value of regression coefficients')
}
\end{lstlisting}  
 
\item 
  Source \texttt{viz.R} to make \texttt{plot\_fitted\_models.R} visible to the global environment.

\item 
  Add the following line in \texttt{analysis.R} and execute it to visualise the estimated coefficients
  of the fitted models (Figure~\ref{fig:fitted_models}):
\begin{lstlisting}
plot_fitted_models(fitted_models)
\end{lstlisting}  
\end{enumerate}

\section{Conclusion}

Although it is not prudent to make general statements based on a single set of simulations, we 
may observe from Figure~\ref{fig:fitted_models} that all three models perform similarly. 
At small samples (N=312), the Normal, Poisson-LogNormal and Negative Binomial models 
correctly identify 7, 8 and 7 eQTLs, respectively, but the Negative Binomial model throws in addition
a false positive. None of these models estimates sufficiently well the size of the eQTLs they 
identify correctly. At higher samples, all models estimate sufficiently well the true eQTLs in the data, 
although they do not avoid the ocassional false positive (see Normal and Poisson-LogNormal models at N=1250
and the Negative Binomial model at N=2500). The Normal model is the fastest to estimate at all sample sizes,
followed by the Poisson-LogNormal model. Proper parameterisation can greatly facilitate inference in
a hierarchical model, and reduce its associated computational cost (see the Stan User's Guide for
 more information).  

Subsequently, we may wish to experiment with additional sparse-inducing priors, and/or 
more systematic and extensive benchmarks, which would require implementing a 
custom inference algorithm for increased performance. This initial analysis using 
a fully automated inference system provides a first assessment of the efficiency of each model, 
and it can serve as a baseline for the subsequent development of novel statistical methods.

\section{Acknowledgments}

This work was supported by the National Institute for Health Research (NIHR) Oxford 
Biomedical Research Centre Program. The views expressed in this manuscript are not 
necessarily those of the Department of Health. We also acknowledge the Wellcome Trust 
Centre for Human Genetics Wellcome Trust Core Award Grant Number 090532/Z/09/Z. The 
funders of the study had no role in the study design, data collection, data analysis, 
data interpretation, or writing of the paper.

\theendnotes{}

\bibliography{chapter}{}
\bibliographystyle{unsrt}

\newpage

\appendix

\section{File \texttt{normal.stan}}

\begin{lstlisting}[language=stan]
data {
  int<lower=0> N;                     // number of genes
  int<lower=0> M;                     // number of samples
  int<lower=0> K;                     // number of variants
  matrix[M, K] X;                     // matrix of genotypes
  vector[M] Y[N];                     // matrix of transformed read counts
}

parameters {
  vector[N] b0;                       // baseline gene expression
  vector[N] ls2;                      // log-variance of gene expression
  real<lower=0> eta;                  // global scale parameter
  vector<lower=0>[K] zeta[N];         // local scale parameters
  vector[K] B[N];                     // regression coefficients
}

transformed parameters {
  vector<lower=0>[N] sigma = sqrt(exp(ls2));     // standard deviation of gene expression
}

model {
  real sc = mean(sigma) / sqrt(N*K);
  eta ~ cauchy(0, 1);
  for(i in 1:N) {
    zeta[i] ~ cauchy(0, 1);
    B[i] ~ normal(0, eta * zeta[i] * sc);
    Y[i] ~ normal(b0[i] + X * B[i], sigma[i]);
  }
}  
\end{lstlisting}

\newpage

\section{File \texttt{poissonln.stan}}

\begin{lstlisting}[language=stan]
data {
  int<lower=0> N;                     // number of genes
  int<lower=0> M;                     // number of samples
  int<lower=0> K;                     // number of variants
  matrix[M, K] X;                     // matrix of genotypes
  int<lower=0> Z[N, M];               // matrix of read counts
  vector<lower=0>[M] c;               // vector of normalisation factors
  vector<lower=0>[M] s;               // vector of library sizes
}

transformed data {
  vector[M] lc = log(c);              // log normalisation factors
  vector[M] ls = log(s);              // log library sizes
}

parameters {
  vector[N] b0;                       // baseline gene expression
  vector[N] ls2;                      // log-variance of gene expression
  real<lower=0> eta;                  // global scale parameter
  vector<lower=0>[K] zeta[N];         // local scale parameters
  vector[K] B[N];                     // regression coefficients
  vector[M] Y[N];                     // latent variables
}

transformed parameters {
  vector<lower=0>[N] sigma = sqrt(exp(ls2));     // standard deviation of gene expression
}

model {
  real sc = mean(sigma) / sqrt(N*K);
  eta ~ cauchy(0, 1);
  for(i in 1:N) {
    zeta[i] ~ cauchy(0, 1);
    B[i] ~ normal(0, eta * zeta[i] * sc);
    Y[i] ~ normal(b0[i] + X * B[i], sigma[i]);
    Z[i] ~ poisson_log(lc + ls + Y[i]);
  }
}  
\end{lstlisting}

\newpage

\section{File \texttt{negbinom.stan}}

\begin{lstlisting}[language=stan]
data {
  int<lower=0> N;                     // number of genes
  int<lower=0> M;                     // number of samples
  int<lower=0> K;                     // number of variants
  matrix[M, K] X;                     // matrix of genotypes
  int<lower=0> Z[N, M];               // matrix of read counts
  vector<lower=0>[M] c;               // vector of normalisation factors
  vector<lower=0>[M] s;               // vector of library sizes
}

transformed data {
  vector[M] lc = log(c);              // log normalisation factors
  vector[M] ls = log(s);              // log library sizes
}

parameters {
  vector[N] b0;                      // baseline gene expression (log-scale)
  vector[N] lphi;                    // log-dispersion
  real<lower=0> eta;                 // global scale parameter
  vector<lower=0>[K] zeta[N];        // local scale parameters
  vector[K] B[N];                    // regression coefficients
}

transformed parameters {
  vector<lower=0>[N] phi = exp(lphi);         // dispersion
  vector<lower=0>[N] alpha = 1.0 ./ phi;      // inverse-dispersion
}

model {
  real sc = 1.0 / sqrt(N*K);
  eta ~ cauchy(0, 1);
  for(i in 1:N) {
    zeta[i] ~ cauchy(0, 1);
    B[i] ~ normal(0, eta * zeta[i] * sc);
    Z[i] ~ neg_binomial_2_log(ls + lc + b0[i] + X * B[i], alpha[i]);
  }
}  
\end{lstlisting}

\newpage

\section{\texttt{R sessionInfo()}}   % chktex 36

\begin{itemize}\raggedright{}
  \item R version 3.5.1 (2018--07--02), \verb|x86_64-pc-linux-gnu|
  \item Locale: \verb|LC_CTYPE=en_GB.UTF-8|, \verb|LC_NUMERIC=C|, \verb|LC_TIME=en_GB.UTF-8|, \verb|LC_COLLATE=en_GB.UTF-8|, \verb|LC_MONETARY=en_GB.UTF-8|, \verb|LC_MESSAGES=en_GB.UTF-8|, \verb|LC_PAPER=en_GB.UTF-8|, \verb|LC_NAME=C|, \verb|LC_ADDRESS=C|, \verb|LC_TELEPHONE=C|, \verb|LC_MEASUREMENT=en_GB.UTF-8|, \verb|LC_IDENTIFICATION=C|
  \item Running under: \verb|Ubuntu 18.04.1 LTS|
  \item Matrix products: default
  \item BLAS:\@ \verb|/usr/lib/x86_64-linux-gnu/openblas/libblas.so.3|
  \item LAPACK:\@ \verb|/usr/lib/x86_64-linux-gnu/libopenblasp-r0.2.20.so|
  \item Base packages: base, datasets, graphics, grDevices, methods, stats, utils
  \item Other packages: bindrcpp~0.2.2, dplyr~0.7.8, forcats~0.3.0, ggplot2~3.1.0, purrr~0.2.5, readr~1.2.1,
    stringr~1.3.1, tibble~1.4.2, tidyr~0.8.2, tidyverse~1.2.1
  \item Loaded via a namespace (and not attached): assertthat~0.2.0, backports~1.1.2, base64enc~0.1--3,
    bindr~0.1.1, broom~0.5.1, callr~3.0.0, cellranger~1.1.0, cli~1.0.1, codetools~0.2--15, colorspace~1.3--2,
    compiler~3.5.1, cowplot~0.9.3, crayon~1.3.4, doMC~1.3.5, foreach~1.4.4, generics~0.0.2, glue~1.3.0,
    grid~3.5.1, gridExtra~2.3, gtable~0.2.0, haven~2.0.0, hms~0.4.2, httr~1.3.1, inline~0.3.15,
    iterators~1.0.10, jsonlite~1.5, labeling~0.3, lattice~0.20--38, lazyeval~0.2.1, loo~2.0.0, lubridate~1.7.4,
    magrittr~1.5, matrixStats~0.54.0, modelr~0.1.2, munsell~0.5.0, nlme~3.1--137, parallel~3.5.1, pillar~1.3.0,
    pkgbuild~1.0.2, pkgconfig~2.0.2, plyr~1.8.4, prettyunits~1.0.2, processx~3.2.1, ps~1.2.1, R6~2.3.0,
    Rcpp~1.0.0, readxl~1.1.0, reshape2~1.4.3, rlang~0.3.0.1, rstan~2.18.2, rstudioapi~0.8, rvest~0.3.2,
    scales~1.0.0, StanHeaders~2.18.0, stats4~3.5.1, stringi~1.2.4, tidyselect~0.2.5, tools~3.5.1, withr~2.1.2,
    xml2~1.2.0, yaml~2.2.0
\end{itemize}

\end{document}